\documentclass{agn_2000}
\usepackage{graphics}
\begin{document}
\title{Optical spectral variability of PG QSOs}
\author{D. Tr\`evese\inst{1} \and F. Vagnetti\inst{2}}  
\institute{Dipartimento di Fisica, Universit\`a di Roma La Sapienza, Piazzale A. Moro 2, I-00185 Roma, Italy
\and Dipartimento di Fisica, Universit\`a di Roma Tor Vergata, Via della Ricerca Scientifica, I-00133 Roma,
Italy}
\maketitle

\begin{abstract}

In previous studies we have shown that the optical variability
of quasars increases, on average, with redshift. We explained this
dependance in terms of a hardening of the spectrum during bright phases,
coupled with the increase of the rest-frame frequency for increasing redshift. We re-analize now these
correlations on the basis of new light curves of PG quasars, recently  published by the
Wise Observatory group. 

\end{abstract} 

\section{Introduction}

     Although variability plays a key role in constraining the models of the
     central engine of QSOs, little is known about the physical origin of
     luminosity variations.  The most diverse variability mechanisms have
     been proposed in the past, including supernovae explosions (Aretxaga
     et al. 1997), instabilities in the accretion disk (Kawaguchi et al. 1998),
     and gravitational lensing due to intervening matter (Hawkins 1993). So
     far, most of the information about the characteristics of variability
     derives from the statistical analysis of single band light curves of
     magnitude limited QSO samples. The correlation of variability with
     either intrinsic luminosity ($v-L$) or redshift ($v-z$) is affected by the
     strong correlation between luminosity and redshift ($L-z$), present in
     these samples. The results of these analyses also depend on the
     specific variability index adopted, as shown by Giallongo et al. (1991),
     who found a positive v-z correlation through a variability index based
     on the rest-frame structure function, later confirmed by Cristiani et al.
     (1996). The increase of variability with the rest-frame observing
     frequency found by Di Clemente et al. (1996), supports the suggestion
     by Giallongo et al. (1991) that QSOs at high redshift appear more
     variable, on average, since they are observed in a higher rest frame
     frequency. The dependence of variability on frequency is associated
     with the hardening of the spectral energy distribution (SED) during the
     bright phases (Cutri et al. 1985, Kinney et al. 1991, Edelson et al.
     1990). Tr\`evese et al. (1999) have shown that a hardening of the SED
     in the bright phase occurs, on average, in the statistical, magnitude
     limited, sample of AGNs of the SA 57 (Tr\`evese et al. 1989, 1994,
     Bershady et al. 1998). They also found that the slope a and its
     variations associated with the luminosity changes, are consistent on
     average with temperature changes of a black body.

\subsection{Data and analysis}
                                                                             
We present preliminary results of a new analysis of the spectral variability of QSOs in the optical band. It is based
on a data set made available by Giveon et al. 1999, consisting of the light curves of a sample of 42 nearby and
bright QSOs ($z<0.4$, $B<16$ mag) belonging to the Palomar-Green (PG) sample, which were monitored for 7 years,
with a typical sampling interval of 39 days, in the Johnson Cousins B and R bands, with the 1 m telescope of the
Wise Observatory. Here the quantities adopted for the analysis are defined. The results are discussed in Section 2,
together with the relevant figures. 

\subsection{Variability index}

We define the structure function as (see Di Clemente et al. 1996)
$S_1=[\pi/2(\langle|m(t+\tau)-m(t)|\rangle^2-\sigma_n^2)]^{1/2}$, where $m(t)$ is either the $B$ or the $R$ magnitude, $t$ is the
rest-frame time, $\tau$ is the time lag between the observations,
$\sigma_n$ is the relevant r.m.s. noise and the angular brackets indicate the average taken over all the pairs of observations
separated by a time interval $\tau\pm\Delta\tau$.  The value of $\Delta\tau$ is the result of a trade-off between time
resolution and statistical uncertainty.  In the following we define, for each QSO, two {\bf variability indexes} $v_B$ and $v_R$
equal to the relevant structure function in the $B$ and $R$ bands, computed for $\tau=2\pm 0.6$ yr. The average values of $v_R$
and
$v_B$ over the ensemble of 42 QSOs, versus the relevant average rest-frame frequency, are reported in Figure 1, which is adapted
from Di Clemente et al. (1996). 

\subsection{Instantaneous spectral slope}

For each QSO we compute the {\bf instantaneous spectral slope} $\alpha(t)\equiv\log(f_{\nu_B}/f_{\nu_R})/\log(\nu_B/\nu_R)$,
regarding as synchronous the observations within a time lag of 9 hours. Emission line corrections are not considered in the
present preliminary analysis. 

\subsection{Spectral variability parameter}

Considering all possible pairs of observations, separated by a time interval $\tau_{ij}\equiv\tau_i-\tau_j$, we define a spectral
variability indicator: $\beta(\tau)\equiv[\alpha(t+\tau)-\alpha(t)]/[\log L_B(t+\tau)-\log L_B(t)]$ where $L_B(t)$ is the absolute
luminosity in the blue band. In the present analysis we consider, for each object, the {\bf spectral variability parameter}
$\beta_m$ defined as the average of all the $\beta(\tau)$ values for $\tau\le 1000$ days.
                                                                          
\section{Results}
 
\subsection{Variability-frequency correlation ($v-\nu$)} 

Fig. 1 shows the variability indexes corresponding to various frequencies. The stars
represent the indexes $v_R$ and $v_B$ computed in the present analysis. The other points,
taken from Di Clemente et al. 1996, derive from different statistical QSO samples as
indicated. The new points are consistent with the general trend and confirm the value of
the slope $\partial S_1/\partial\log\nu_{rest}=\partial S_1/\partial\log(1+z)=0.25-0.30$ which accounts for the $v-z$
correlation found by Giallongo et al. (1991). A $v-z$ correlation is not detectable for the
whole sample (Giveon et al. 1999), due to the small redshift range ($z<0.4$). If we restrict
to an absolute magnitude bin $-23.5<M_B<-22.5$, to reduce the combined effect of the
(positive) $L-z$ and (negative) $v-L$ correlations, which attenuates the apparent the $v-z$
correlation, we find a partial correlation coefficient $r_{v,z}=0.39$, marginally significant
[$P(>r)=0.09$], despite the small number (19) of QSOs in the selected magnitude bin.
Therefore, we can conclude that the intrinsic $v-z$ correlation is present, and the trend
shown in Fig. 1 confirms its interpretation in terms of an increase of variability for
increasing rest-frame frequency. 

\begin{figure}
 \resizebox{\hsize}{!}{\includegraphics{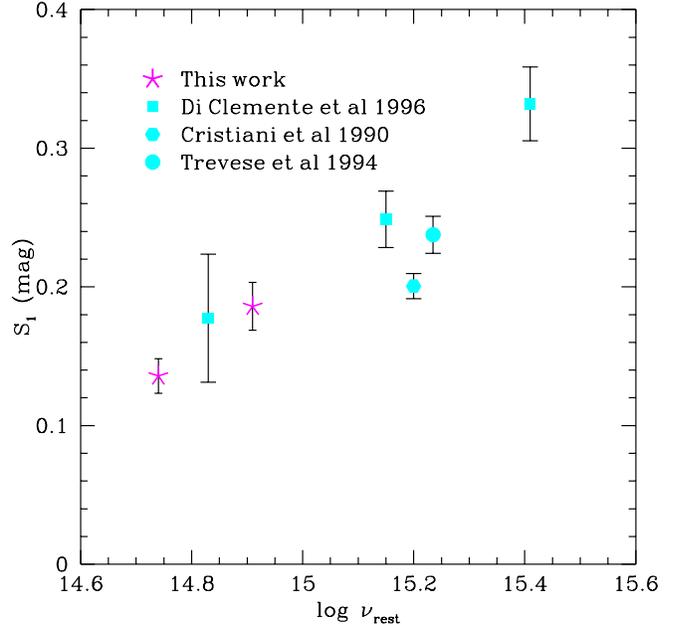}}
\caption[]{Variability indexes as a function of rest-frame frequency adapted from Di Clemente et al. 1996.}
\end{figure}

\subsection{Spectral slope - luminosity ($\alpha-L_B$) correlation} 

In Fig. 2 the instantaneous slope $\alpha(t)$ of each QSO is plotted as a function of the
relevant blue luminosity $L_B$. The regression lines $\alpha$ vs. $L_B$ are also reported for each QSO. 
Two features appear in the figure:
(i) {\bf Intra-QSO correlation ($\Delta\alpha - \Delta\log L$)}, i.e. a trend (with a few exceptions) of
increasing $\alpha$ for increasing $\log L_B$ appears, indicating that QSO SEDs are harder in the
bright phase (and vice versa);
(ii) {\bf Inter-QSO correlation ($\alpha-\log L$)}, i.e. also the average $\alpha$ of each QSO is positively
correlated with its average luminosity.

\begin{figure}
 \resizebox{\hsize}{!}{\includegraphics{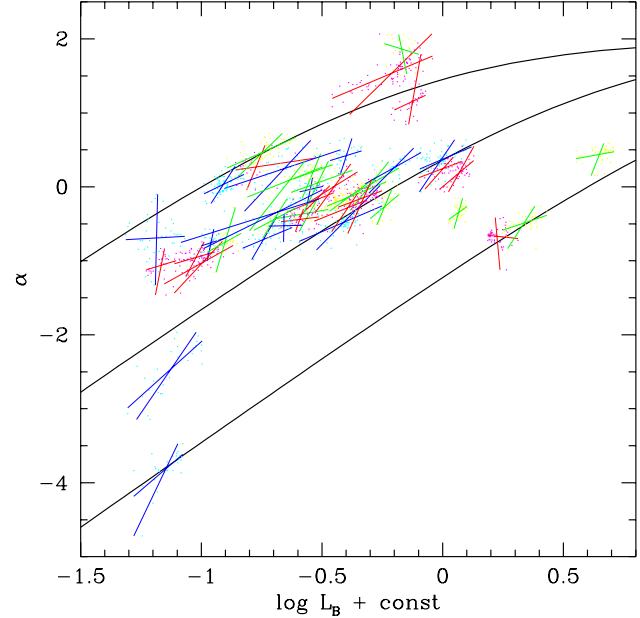}}
\caption[]{Instantaneous slope $\alpha(t)$ of each QSO, plotted as a function of the
relevant blue luminosity $L_B$. The regression lines $\alpha$ vs. $\log L_B$ are also reported for each QSO.}
\end{figure}

These two effects suggest that both the change of $\alpha$
during luminosity variations and the average increase of $\alpha$ from faint to bright
QSOs have the same  physical origin: the increase of the temperature of the
emitting gas. As an example the $\alpha(t)$ versus $\log L_B(t)$ lines computed for three
emitting black bodies of different areas are reported in Fig. 2. The temperature
increases along each line from bottom left to top right. 

\subsection{Spectral variability versus spectral slope $\beta-\alpha$} 

Fig. 3 shows the spectral variability parameter $\beta_m$ of each QSO, versus its average
spectral slope $\alpha_m$. The continuous curve represents $\beta$ versus $\alpha$ for a family of black
bodies of different temperatures, increasing from left to right. This is not a fit to the data
points, since there are no free parameters in the curve. Comparison with the data suggests
that the spectral slope and the slope variations are roughly consistent with a black body of
tempeature $T=2.5\cdot 10^4$ K, subject to small temperature changes responsible of both the
luminosity and spectral slope changes. 

\begin{figure}
 \resizebox{\hsize}{!}{\includegraphics{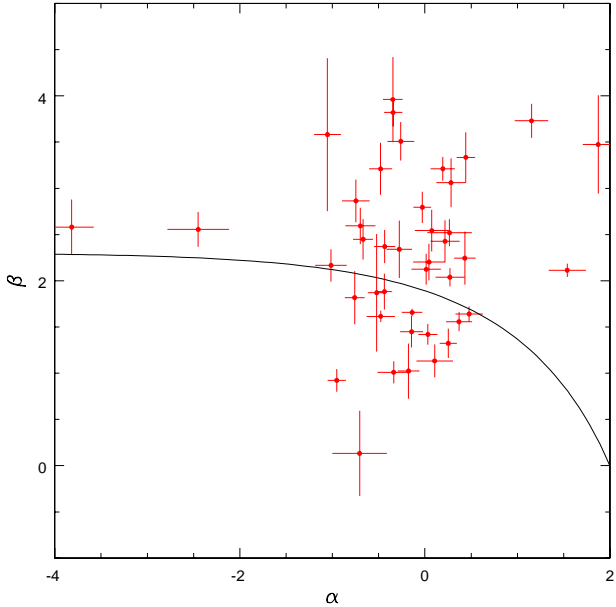}}
\caption[]{Spectral variability parameter versus spectral slope.}
\end{figure}

\section{Conclusions}

The analysis of the spectral variability of the present statistical subsample of PG QSOs
shows that: 
\begin{itemize}
\item
the amplitude of variability increases with the rest-frame frequency; 
\item
a (positive) $v-z$ correlation  appears in the present sample,  despite the
(negative) $v-L$ and the (positive) $L-z$ correlations; 
\item
the increase of variability with frequency is consistent with the trend
previously found by Di Clemente et al. (1996) on the basis of different QSO
samples and confirms the explanation of the $v-z$ correlation suggested by
Giallongo et al. (1991); 
\item
the spectral slope increases on average in the bright phases (intra-QSO $\alpha-L$
correlation); 
\item
the spectral slope is steeper, on average, for fainter QSOs (inter-QSO $\alpha-L$
correlation); 
\item
the two latter points suggest that the gas temperature is  responsible for
both the effects; 
\item
temperature variations of a black body with an average temperature of the
order of $2.5\cdot 10^4$ K are consistent for the intra-QSO $\alpha(t) - L_B(t)$ correlation.  
\item
different average black body temperatures may account for the inter-QSO
$\alpha - L_B$ correlation. 
\end{itemize}

\begin{acknowledgements}
We are grateful to the Wise Observatory Group for providing public access to their data.
\end{acknowledgements}


\begin{thebibliography}{}
\bibitem{} Aretxaga, I., Cid Fernandes, R., Terlevich, R.J. 1997, MNRAS 286, 271 
\bibitem{} Bershady, M. A., Tr\`evese, D., Kron, R.G. 1997, ApJ 496, 103 
\bibitem{} Cristiani, S., Trentini, S., La Franca, F., Aretxaga, I., Andreani, P., Vio, R., Gemmo, A. 1996, A\&A 306, 395
\bibitem{} Cristiani, S., Vio, R., Andreani, P. 1990, AJ 100, 56  
\bibitem{} Cutri, R.M., Wisniewski, W.Z., Rieke, G.H., Lebofsky, H.J. 1985, ApJ 296, 423 
\bibitem{} Di Clemente, A., Giallongo, E., Natali, G., Tr\`evese, D., Vagnetti. F. 1996, ApJ 463, 466 
\bibitem{} Edelson, R.A., Krolik, J. H., Pike, G. F. 1990, ApJ 359, 86
\bibitem{} Giallongo, E., Tr\`evese, D., Vagnetti, F. 1991, ApJ 377, 345 
\bibitem{} Giveon, U., Maoz, D., Kaspi, S., Netzer, H., Smith, P.S. 1999, MNRAS 306, 637 
\bibitem{} Kawaguchi, T., Mineshige, S., Umemura, M., Turner, E.L. 1998, ApJ 504, 671 
\bibitem{} Kinney, A. L., Bohlin, R.C., Blades, J.C., York, D.G. 1991, ApJSupp 75, 645 
\bibitem{} Hawkins, M. R. S. 1993, nature 366, 242 
\bibitem{} Tr\`evese, D., Kron, R.G., Majewski, S.R., Bershady, M.A., Koo, D.C. 1994 ApJ 433, 494 
\bibitem{} Tr\`evese D., Pittella, G., Kron, R.G., Koo, D.C., Bershady, M.A. 1989, AJ 98, 108 
\bibitem{} Tr\`evese D., Bunone, A., Kron R. G. 1999, Mem SAIt 70, 37
\end{thebibliography}
\end{document}